\documentclass{DISproc}

\begin{document}
\title{ABM11 parton distributions and benchmarks}
\author{{\slshape Sergey Alekhin$^{1,2}$, Johannes Bl\"umlein$^1$,
Sven-Olaf Moch$^1$}\\[1ex]
$^1$DESY, Platanenallee 6, D--15738 Zeuthen, Germany\\
$^2$Institute for High Energy Physics,
    142281 Protvino, Moscow region, Russia}

\contribID{302}

\doi  

\maketitle

\vspace*{-55mm}
DESY-12-139\\
$~~~~~$LPN12-086\\
$~~~~~$SFB/CPP-12-53
\vspace*{40mm}

\begin{abstract}
We present a determination of the nucleon parton
distribution functions (PDFs) and of the strong coupling constant $\alpha_s$
at next-to-next-to-leading order (NNLO) in QCD
based on the world data for deep-inelastic scattering and the fixed-target data
for the Drell-Yan process.
The analysis is performed in the fixed-flavor number scheme for $n_f=3,4,5$
and uses the $\overline{MS}$ scheme for $\alpha_s$ and the heavy quark masses.
The fit results are compared with other PDFs and used to
compute the benchmark cross sections at hadron colliders to the NNLO accuracy. 
\end{abstract}

\noindent
The nucleon PDFs play crucial role in the collider phenomenology and very 
often they put a limit on theoretical prediction accuracy, particularly for the 
calculations in the next-to
\begin{wrapfigure}{r}{0.45\textwidth}
  \centering
  \includegraphics[width=0.4\textwidth]{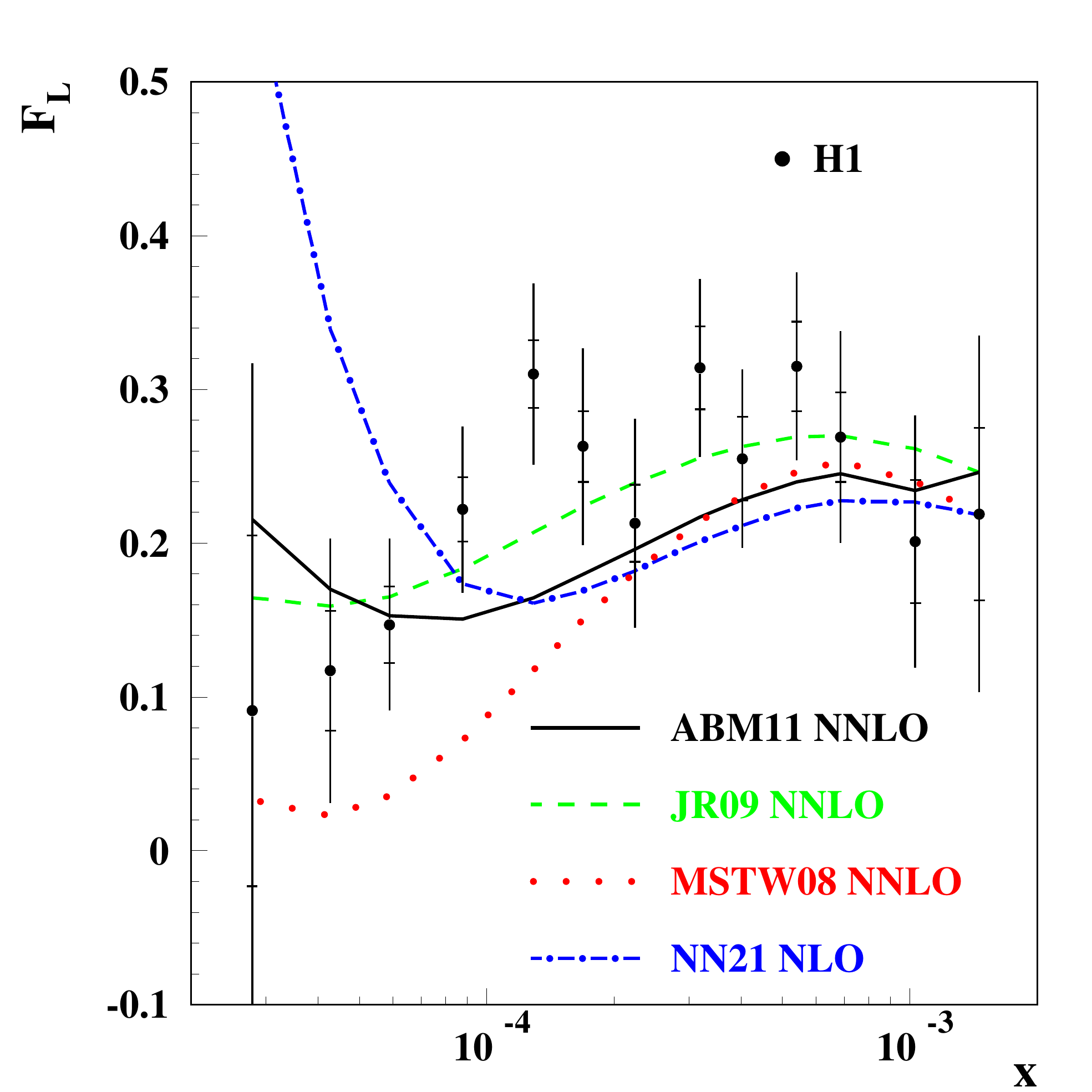}
  \caption{\small      The data on $F_L$ versus $x$ obtained by the H1
     collaboration~\cite{Aaron:2010ry} 
     confronted with the 3-flavor scheme NNLO predictions based on the 
     different PDFs (solid line: this analysis, dashes: 
     JR09~\cite{JimenezDelgado:2008hf}, dots: MSTW~\cite{Martin:2009iq}). 
     The NLO predictions based on the 3-flavor NN21 
     PDFs~\cite{Ball:2011mu} are given for comparison (dashed dots). 
     The value of $Q^2$ for the data points and the curves in the plot 
     rises with $x$ in the range of $1.5 \div 45~{\rm GeV}^2$.}
  \label{fig:fl}
\end{wrapfigure}
-next-to-leading order (NNLO) in QCD.
To meet quick accumulation of the data and steady progress 
in reduction of the systematic uncertainties in 
the LHC experiment we provide the NNLO nucleon PDF set with improved accuracy~\cite{Alekhin:2012ig}. 
These PDFs are 
obtained from the updated version of the ABKM09 analysis~\cite{Alekhin:2009ni}
 performed in the fixed-flavor number (FFN) scheme with the number of 
fermions taken as $n_f=3,4,5$, depending on the process used to constrain 
the PDFs. In the present analysis we replace the inclusive neutral-current 
(NC) DIS data of the H1 and ZEUS experiments by the combined HERA data set, which are
obtained from merging those of separate experiments~\cite{Aaron:2009wt}. The data are 
substantially improved by cross-calibration of the separate experiments and 
by  merging both statistical and systematic errors. Due to these improvements the 
combined HERA data provide a better constraint
on the small-$x$ gluon and quark distributions. We also add to our analysis 
the inclusive charged-current (CC) DIS HERA data obtained by merging 
the H1 and ZEUS samples. 
The CC HERA data provide a supplementary constraint 
on the PDFs helping to disentangle the small-$x$ quark distributions. Finally, 
we include the H1 data obtained in a special HERA run at 
reduced collision energy, which are particularly sensitive to the 
contribution of longitudinal structure function $F_L$ at small 
$x$~\cite{Aaron:2010ry}.  
This run was motivated by a particular sensitivity of the small-$x$ $F_L$ 
to the resummation effects and collinear factorization violation. 
Besides, $F_L$ is quite sensitive to the gluon distribution therefore 
the data of Ref.~~\cite{Aaron:2010ry} can help to consolidate 
the small-$x$ gluon distributions provided by different groups, 
cf. Fig.~\ref{fig:fl}.

\begin{figure}[htb]
  \centering
  \includegraphics[width=0.45\textwidth]{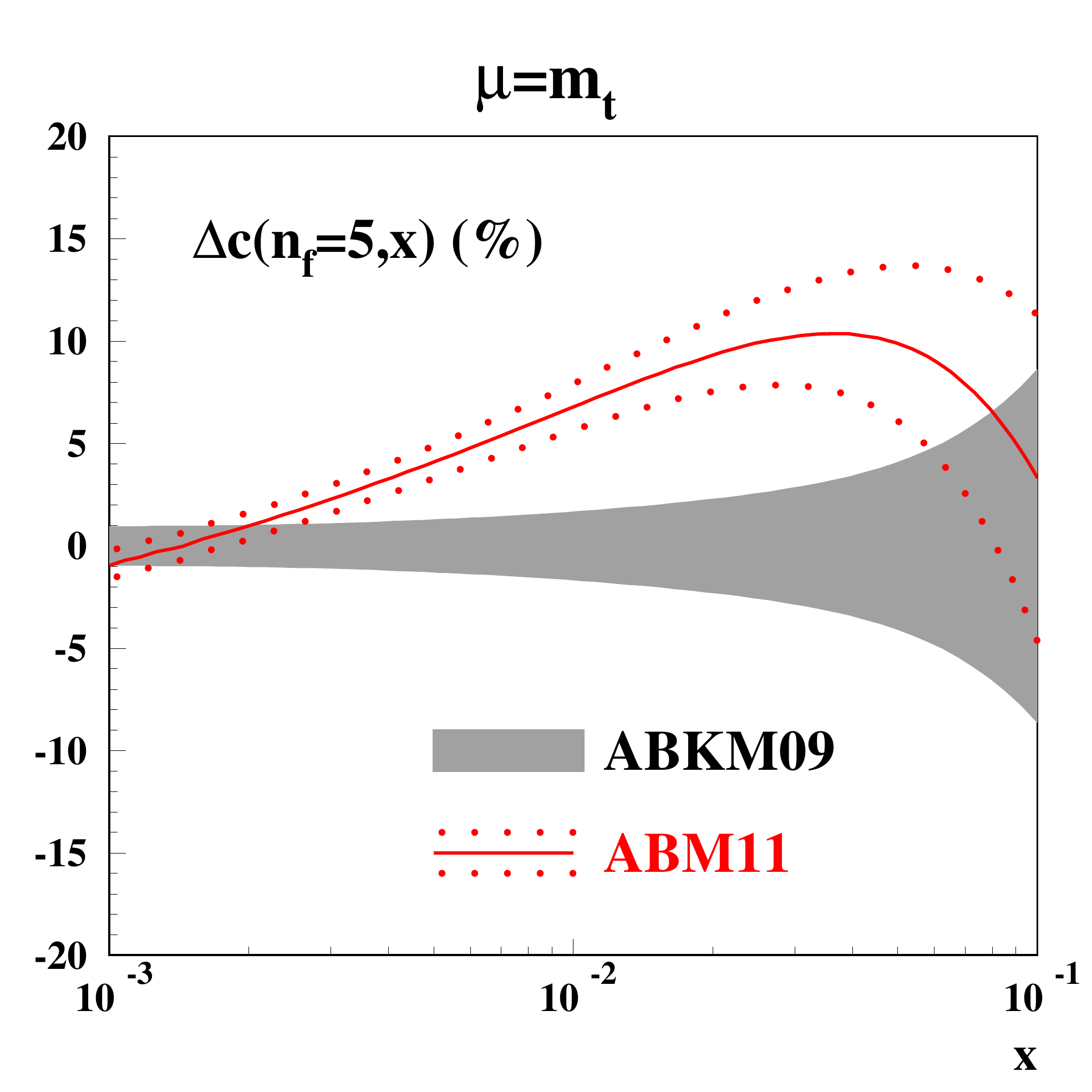}
  \includegraphics[width=0.45\textwidth]{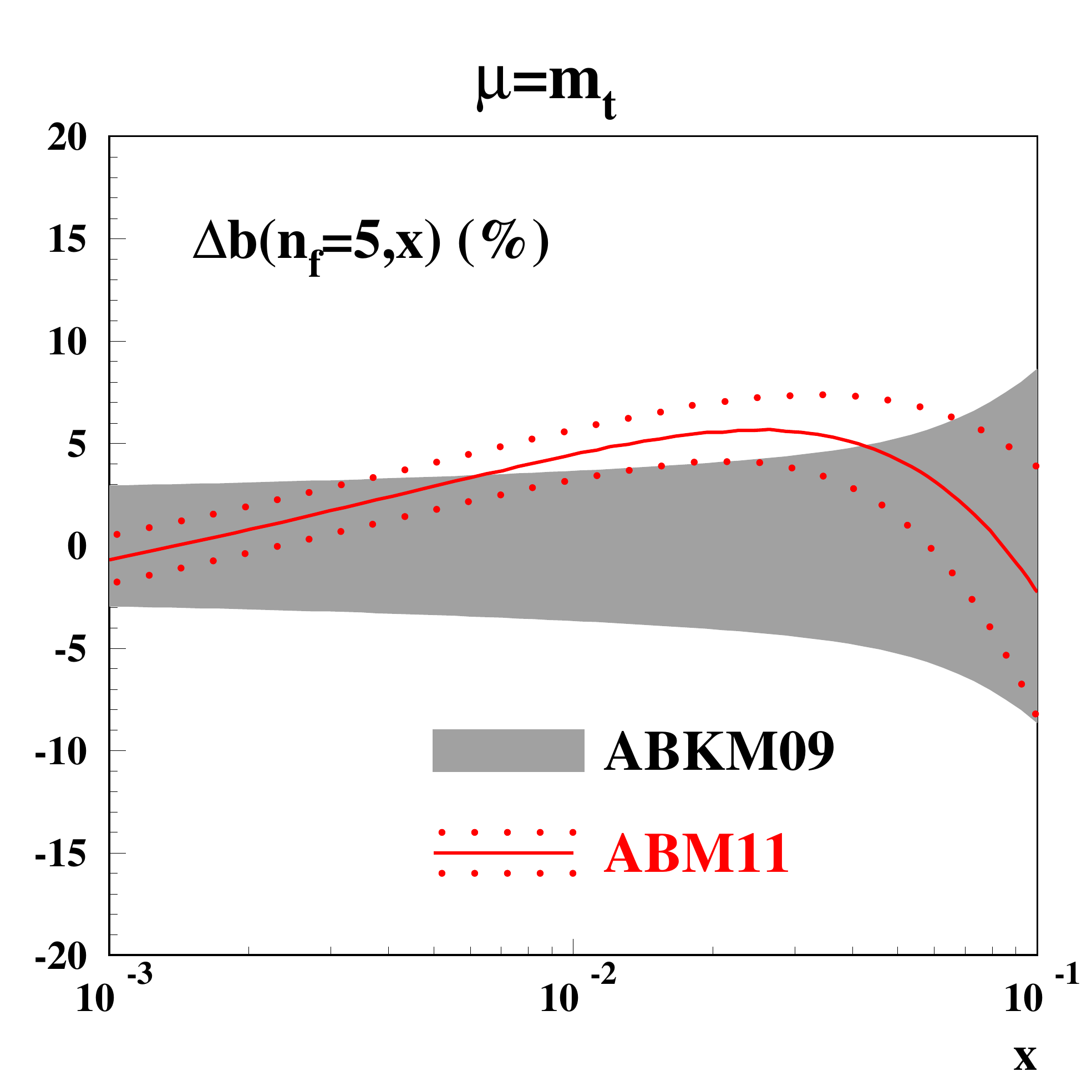}
  \caption{\small The charm- (left) and the bottom-quark (right) 
      PDFs obtained in the fit:
      The dotted (red) lines denote the $\pm 1 \sigma$ band of relative
      uncertainties (in percent) and the solid (red) line indicates the
      central prediction resulting from the fit with
      the running masses taken at the PDG values~\cite{Nakamura:2010zzi}.
      For comparison the shaded (grey) area represents the 
      results of ABKM09~\cite{Alekhin:2009ni}.
}
  \label{fig:hq}
\end{figure}

In our analysis the DIS data are described within the 3-flavour 
FFN scheme, as well as in the ABKM09 case. However, in the present fit 
we employ the heavy-quark Wilson coefficients with the $\overline{MS}$ 
definition for the $c$- and $b$-quark masses, as suggested in 
Ref.~\cite{Alekhin:2010sv}. 
For the case of $\overline{MS}$ definition the perturbative stability 
of the calculations is substantially improved. Moreover, in this case
the constraints  
on the heavy-quark masses coming from the $e^+e^-$ data, which are commonly   
obtained in the $\overline{MS}$ definition, can be consistently imposed 
in the PDF fit. This leads to a reduction of the PDF uncertainties 
due to the heavy-quark masses. In particular, the errors in the 
4(5)-flavour heavy-quark PDFs, which are  generated from the 3-flavour ones 
using the matching conditions, are significantly improved as compared to 
the earlier ABKM09 PDFs, cf. Fig.~\ref{fig:hq}.

\begin{figure}[htb]
  \centering
  \includegraphics[width=\textwidth,height=8cm]{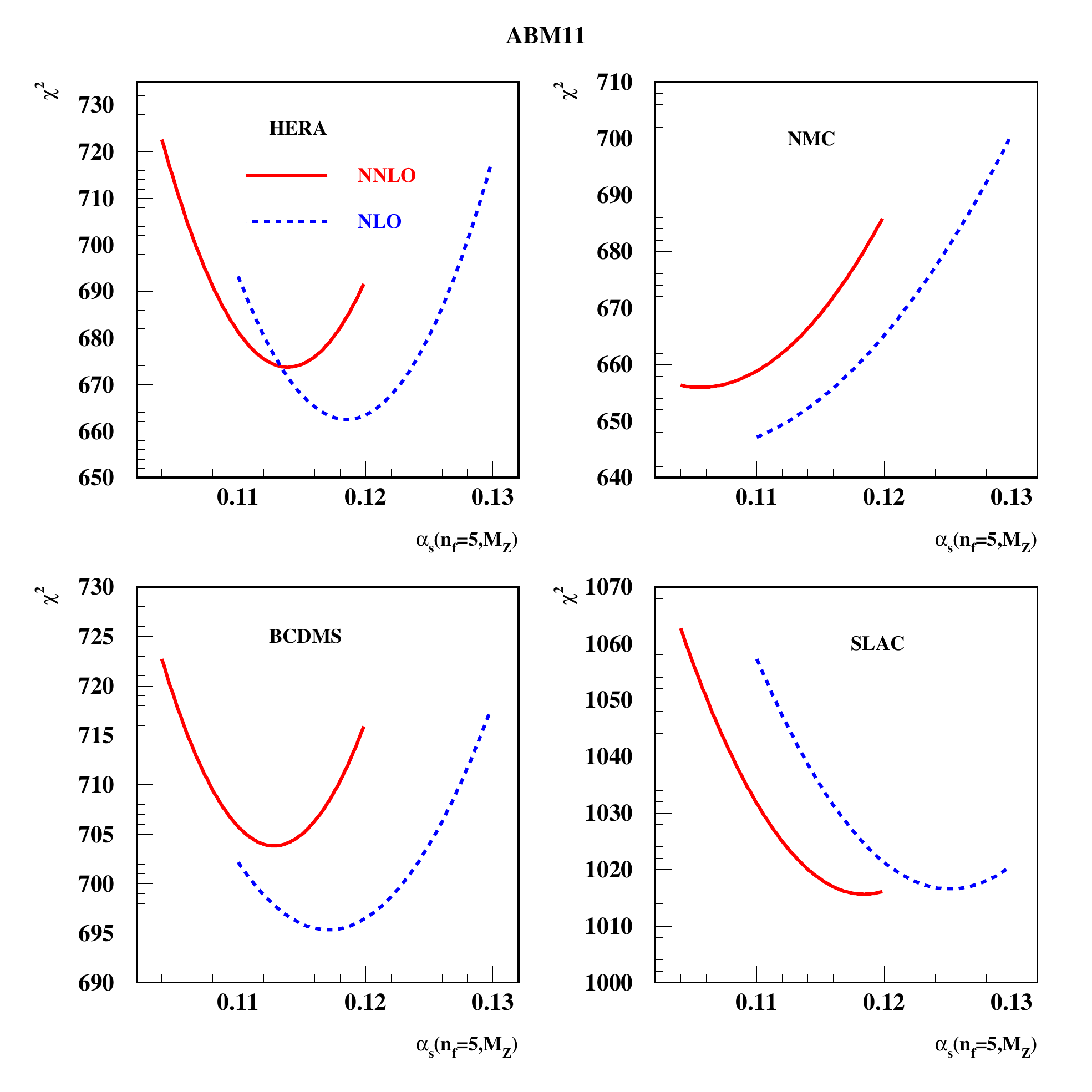}
  \caption{\small 
    The $\chi^2$-profile versus the value of $\alpha_s(M_Z)$, 
        for the separate data subsets,  
        all obtained in variants of the present analysis with the 
       value of $\alpha_s$ fixed 
    and all other parameters fitted
    (solid lines: NNLO fit, dashes: NLO fit).
}
  \label{fig:alp}
\end{figure}

The value of strong coupling constant $\alpha_s(M_Z)$ is determined 
in our fit 
simultaneously with the PDFs. This approach provides a straightforward
treatment of their correlation that is important for calculation 
of the uncertainties in the hadronic cross section predictions.  
At NNLO the ABM11 fit obtains the value of 
$\alpha_s(M_Z)=0.1134\pm0.0011({\rm exp.})$. 
This is comparable with our earlier determination 
$\alpha_s(M_Z)=0.1135\pm0.0014({\rm exp.})$~\cite{Alekhin:2009ni}, while the 
error is improved due to more accurate data employed in the present analysis.   
It is also in a good agreement with 
$\alpha_s(M_Z)=0.1141^{+0.0020}_{-0.0022}$
obtained in the analysis of the non-singlet DIS data with account of the QCD 
corrections up to the N$^3$LO~\cite{Blumlein:2006be}.
In the ABM11 analysis the value of $\alpha_s$ is constrained both by the 
non-singlet and the singlet DIS data, cf. Fig.~\ref{fig:alp}.
For the kinematics of the SLAC and NMC experiments
the $\chi^2$-profile is sensitive to the power corrections including
target mass effects and the dynamical twist-4 terms. The latter are 
poorly defined by the strong interaction theory and therfore put 
limit on the 
accuracy of $\alpha_s$ determined in our fit. On the other hand, the BCDMS and HERA 
data are insensitive to the power term due to kinematics peculiarities. 
Moreover, these data sets provide complementary constraints in 
determining $\alpha_s$~\cite{Adloff:2000qk}. Performing the NNLO variant of our fit 
with the SLAC and NMC data dropped we obtain 
$\alpha_s(M_Z)=0.1133\pm0.0011({\rm exp.})$, which is not affected
by the power corrections. Furthermore, it is 
in nice agreement with one obtained in the 
nominal ABM11 fit that gives confidence in the consistent treatment 
of the power terms in our analysis (cf. also discussion in 
Ref.~\cite{Alekhin:2011ey}).

\begin{figure}[th!]
  \centerline{
    \includegraphics[width=0.45\textwidth,height=5cm]{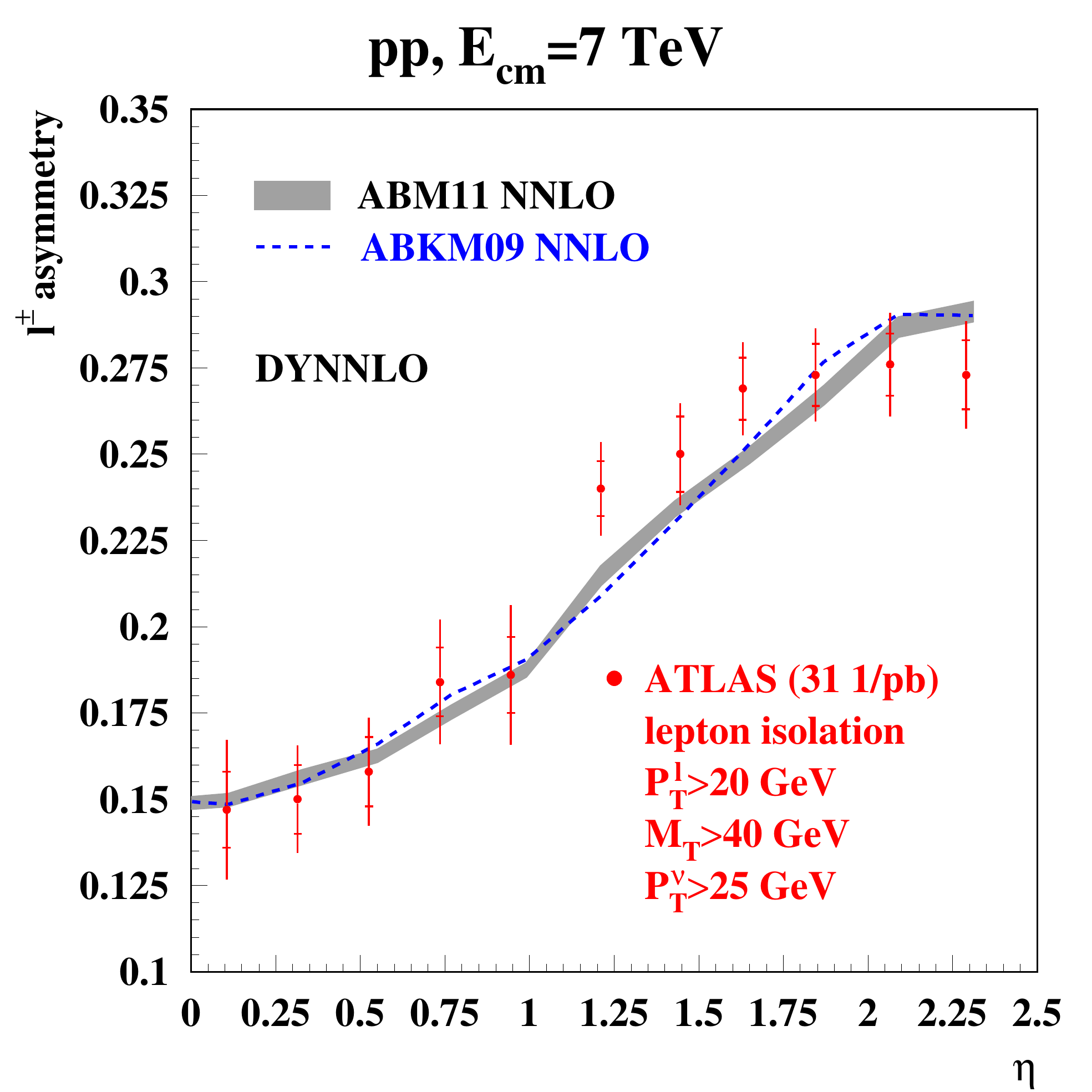}
    \includegraphics[width=0.45\textwidth,height=5cm]{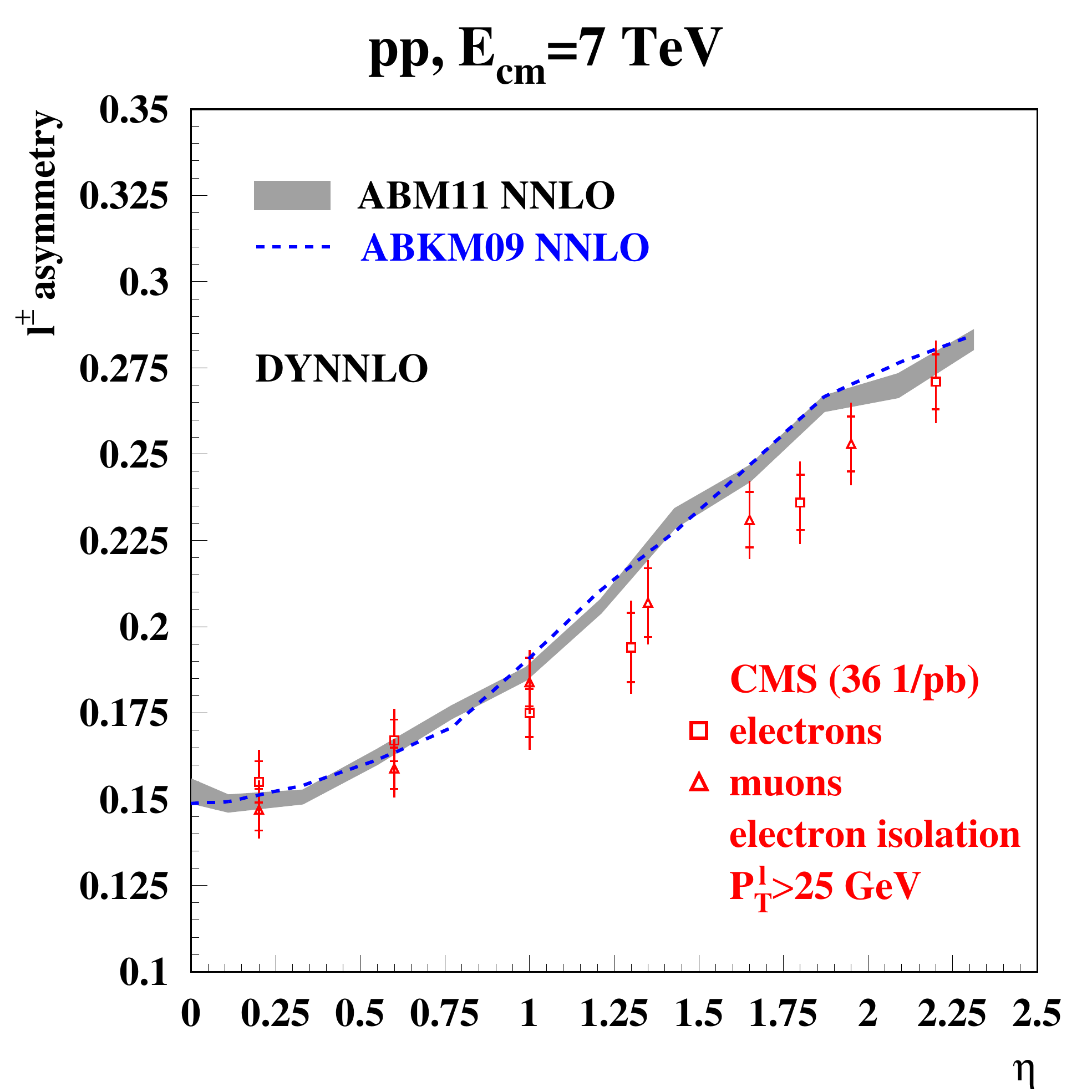}}
  \caption{\small
    \label{fig:was}
     The data on charged-lepton asymmetry versus the 
     lepton pseudo-rapidity $\eta$
     obtained by the ATLAS~\cite{Aad:2011yn} (left panel)
     and CMS~\cite{Chatrchyan:2011jz} (right panel) experiments 
     compared to the NNLO predictions based on the 
     {\tt DYNNLO} code~\cite{Catani:2009sm} and 
     the ABM11 NNLO PDFs with the shaded area showing the integration uncertainties. 
     The ABKM09 NNLO predictions are given for comparison by dashes, 
     without the integration uncertainties shown. 
  }
\end{figure}

\begin{figure}[htb]
  \centering
  \includegraphics[width=0.45\textwidth]{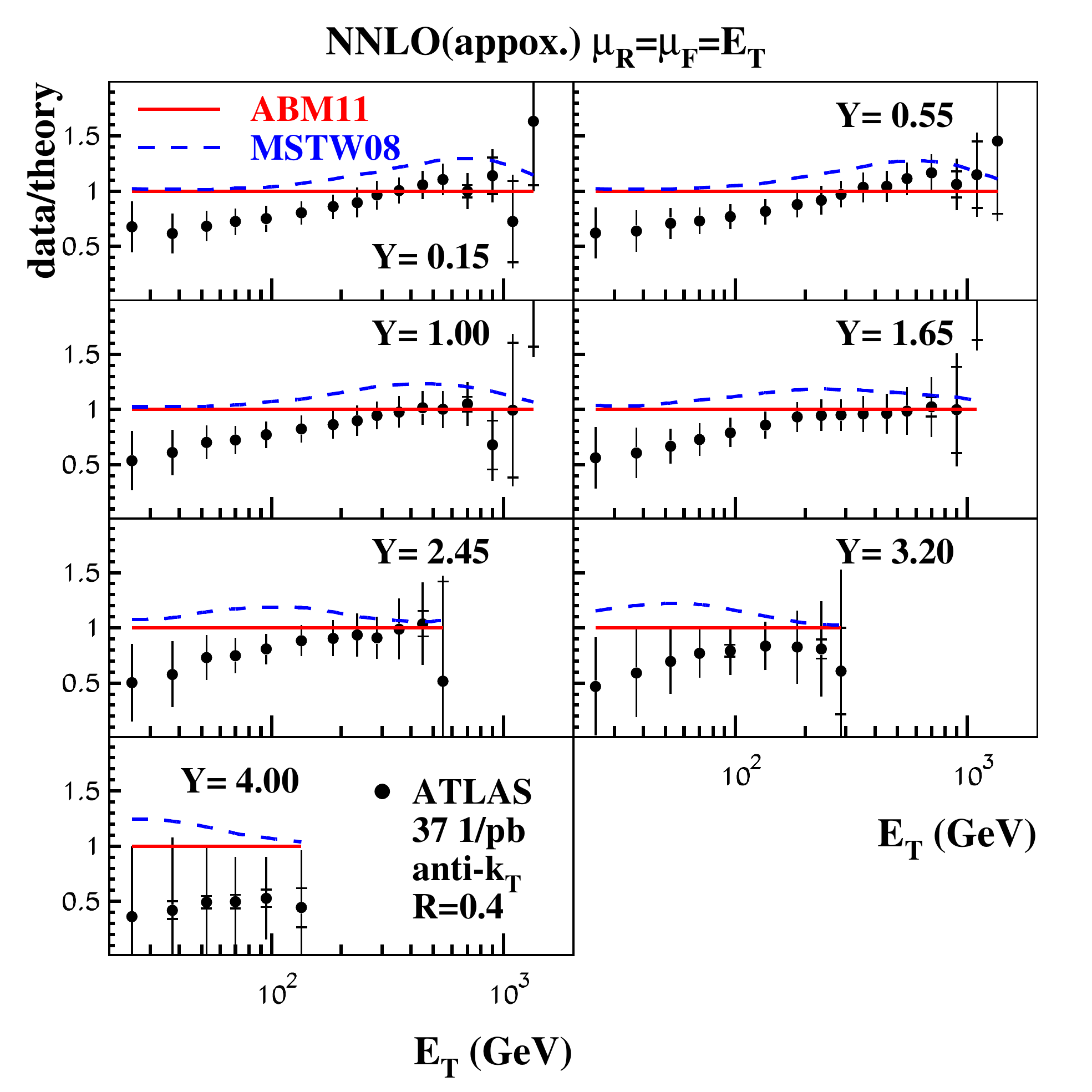}
  \includegraphics[width=0.45\textwidth]{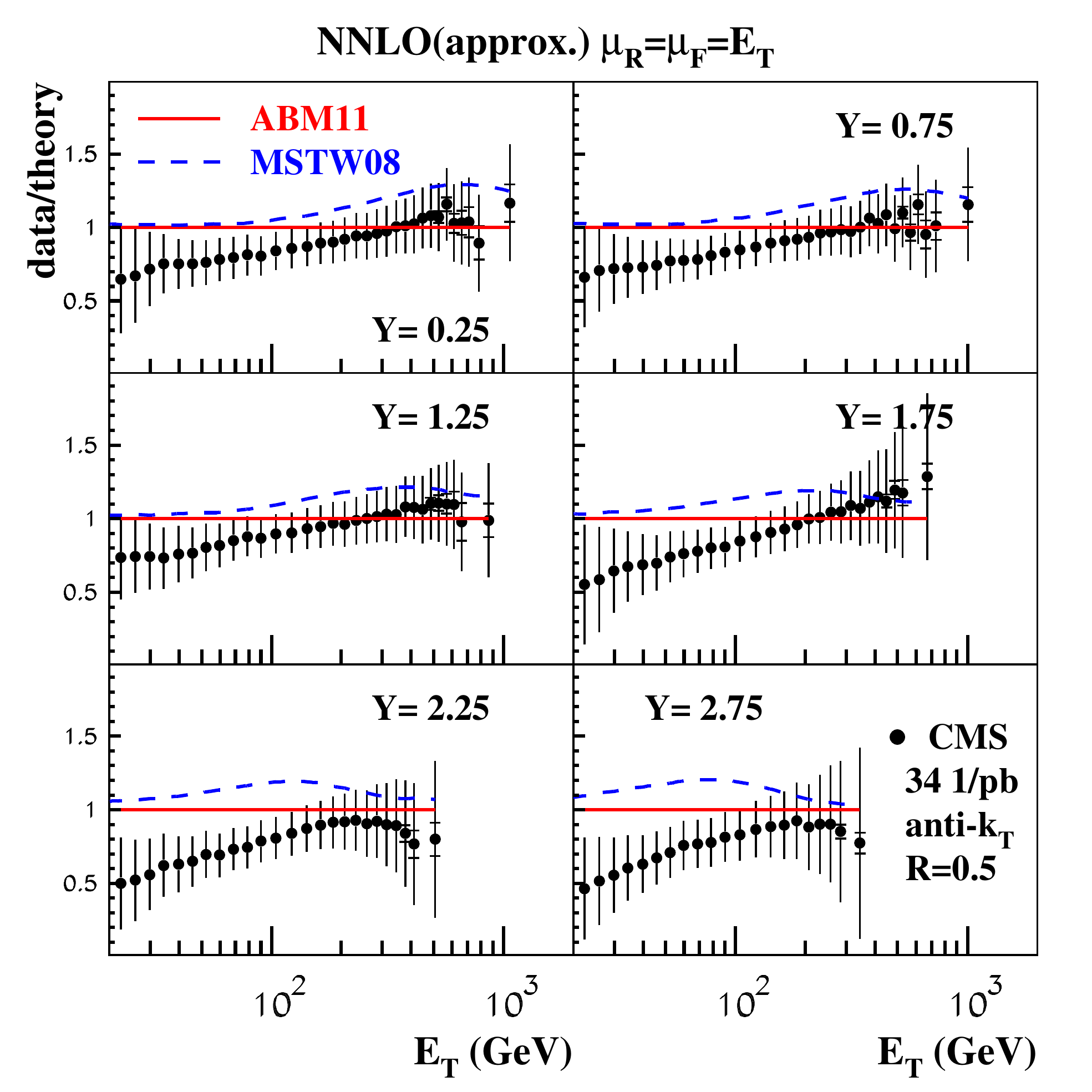}
  \caption{Cross section data for 1-jet inclusive production from 
the ATLAS collaboration~\cite{Aad:2011fc} (left panel) and from the CMS 
collaboration~\cite{CMS:2011ab} (right panel) as a
function of the jet’s transverse energy ET for $\mu_R= \mu_F = E_T$ compared 
to the result of the present analysis
(solid) and to MSTW~\cite{Martin:2009iq} (dashed). 
The theory predictions include 
the NNLO threshold resummation corrections to the jet production.
  \label{fig:jet}}
\end{figure}

Predictions for the charged-lepton asymmetry and the 
inclusive jet production cross sections at the 
energy of LHC are in a good agreement with the first data collected by
the CMS and ATLAS 
experiments~\cite{Aad:2011yn,Chatrchyan:2011jz,Aad:2011fc,CMS:2011ab}, 
cf. Figs.~\ref{fig:jet},~\ref{fig:was}, despite these
data are not used in ABM11 fit. Moreover, 
the value of $\alpha_s=0.1151\pm0.0001~({\rm stata.})\pm0.0047({\rm sys.})$ 
extracted from the ATLAS data of Ref.~\cite{Aad:2011fc} in the 
NLO~\cite{Malaescu:2012ts} is in agreement with our results. 
In contrast, the jet Tevatron data go above 
our predictions and the large-$x$ gluon distribution rises significantly 
once they are included in the analysis. 
Note that the MSTW PDFs systematically overshoot the 
LHC jet data (cf. Fig.~\ref{fig:jet}) as well as other PDFs 
tuned to the Tevatron data~\cite{CMS:2011ab,Aad:2011fc}.
On the whole, this leads to the conclusion that the LHC data 
prefer softer gluons as compared to the Tevatron case. 

The Higgs production rates at the LHC and Tevatron are widely 
defined by the gluon distribution shape and the value of $\alpha_s$. 
The NNLO predictions for the cross section of Higgs production in the 
proton-proton collisions at the LHC energies calculated with different 
NNLO PDFs are displayed in Table~\ref{tab:higgs}.
At smaller collision energy, when the production rate is more sensitive 
to the large-$x$ gluon distribution tail, the ABM11 calculations go lower 
than the MSTW08 and NN21 ones, while at high energies the difference between 
the predictions is smaller. 
\begin{table}[th!]
\renewcommand{\arraystretch}{1.3}
\begin{center}
{\small
\begin{tabular}{|c|c|c|c|c|c|}
\hline
{$\sqrt s$}
&{ABM11}
&{ABKM09~\cite{Alekhin:2009ni}}
&{JR09~\cite{JimenezDelgado:2008hf,JimenezDelgado:2009tv}}
&{MSTW08~\cite{Martin:2009iq}}
&{NN21~\cite{Ball:2011uy}}
\\     
{(TeV)}
&
&
&
&
&
\\     
\hline
 7
 &$ 13.23^{+ 1.35}_{- 1.31}{^{+ 0.30}_{- 0.30}}$
 &$ 13.12^{+ 1.34}_{- 1.31}{^{+ 0.38}_{- 0.38}}$
 &$ 13.02^{+ 1.24}_{- 1.17}{^{+ 0.41}_{- 0.41}}$
 &$ 14.39^{+ 1.54}_{- 1.47}{^{+ 0.17}_{- 0.22}}$
 &$ 15.14^{+ 1.68}_{- 1.53}{^{+ 0.21}_{- 0.21}}$
 \\
8
 &$ 16.99^{+ 1.69}_{- 1.63}{^{+ 0.37}_{- 0.37}}$
 &$ 16.87^{+ 1.68}_{- 1.63}{^{+ 0.47}_{- 0.47}}$
 &$ 16.53^{+ 1.54}_{- 1.44}{^{+ 0.53}_{- 0.53}}$
 &$ 18.36^{+ 1.92}_{- 1.82}{^{+ 0.21}_{- 0.28}}$
 &$ 19.30^{+ 2.09}_{- 1.89}{^{+ 0.26}_{- 0.26}}$
 \\
14
 &$ 44.68^{+ 4.02}_{- 3.78}{^{+ 0.85}_{- 0.85}}$
 &$ 44.75^{+ 4.07}_{- 3.85}{^{+ 1.16}_{- 1.16}}$
 &$ 42.13^{+ 3.60}_{- 3.26}{^{+ 1.59}_{- 1.59}}$
 &$ 47.47^{+ 4.52}_{- 4.18}{^{+ 0.50}_{- 0.71}}$
 &$ 49.77^{+ 4.91}_{- 4.30}{^{+ 0.54}_{- 0.54}}$
 \\
\hline
\end{tabular}
\label{tab:higgs}
}
\caption{\small 
The total NNLO cross sections in pb for Higgs production 
in the gluon-gluon fusion obtained with 
different PDF sets at the mass of Higgs boson $M_H=125~{\rm GeV}$. 
The errors shown are the scale uncertainty are based 
on the shifts $\mu=m_H/2$ and $\mu = 2m_H$ 
and the 1$\sigma$ PDF uncertainty, respectively.  
}
\end{center}
\end{table}

In summary, we have produced the new NNLO PDF set with improved accuracy at 
small $x$ due to new input from the HERA data and refined theoretical 
treatment of the heavy-quark electro-production in the running-mass 
definition. The predictions based on these PDFs are in a good agreement with 
the first LHC data, which can be used in future to improve the PDF
accuracy further. A benchmarking w.r.t. to other
PDFs is performed; the differences found can be also reduced
with the help of new HERA and LHC data.

\begin{footnotesize}
\bibliographystyle{DISproc}
\bibliography{alekhin_sergey.bib}
\end{footnotesize}

\end{document}